# Analytical Equations to the Chromaticity Cone: Algebraic Methods for Describing Color

Prashanth Alluvada
Department of Electrical Engineering
University of Arkansas, Fayetteville
Email: palluvad@uark.edu

**Abstract**

We describe an affine transformation on the (CIE) color matching functions and map the spectral locus as a circle. We then homogenize the right circular cylinder erected by the circle, with respect to a normalizing plane and develop an analytical equation to the chromaticity cone, for the spectral colors. In the interior of the (CIE) chromaticity diagram, by homogenizing elliptic cylinders with respect to the normalizing planes, analytical equations to subsets (also cones) of the chromaticity cone are developed. These equations provide an algebraic method for describing color perception. As an application of the interior chromaticity cones, we demonstrate that by sectioning homogenized cones with planes and projecting, analytical equations to the Macadam ellipses may be derived. Further, the cone equations are used to propose new types of color order systems.

## 1. Introduction

Under the assumptions of color matching experiment, the color matching functions (CMF's), determine a conical region called chromaticity cone (CC), which contains all the excitation levels of the three cones. Since all spectral radiance distributions (SRD) cause the cones to excite to values within or on the CC, the conical region fully describes the color perception of the human eye. The boundary of the CC may be obtained by running a monochromatic SRD of varying heights through the visible range and plotting the cone excitation levels (tri-stimulus values) in a suitably defined coordinate system (Eqn. 2.1a-c). Through affine transformations, homogenizing transformations and employing the principles of Analytical Geometry we obtain full description of the chromaticity cone through algebraic expressions. As an application of the expressions, we demonstrate the mechanism through which Macadam ellipses occur on the CIE chromaticity diagram.

## 2. The Affine-Homogeneous Transformation

With the usual notation, $x(\lambda), y(\lambda)$ and $z(\lambda)$ represent the CIE standard observer CMF's. For the given SRD, the tri-stimulus values (X,Y,Z) and the chromaticity coordinates (x,y, Eqn. (2.1d,e)) are computed using their defining equations Eqn. (2.1). $L_{\varepsilon,\lambda}$ is the SRD. By sweeping a single spike (monochromatic SRD) of unit height through the visible range of the electromagnetic spectrum and using the definitions ([1], [2], Eqn. (2.1)), the spectral locus of the chromaticity diagram may be constructed. However, since the analytical equation of this curve is unknown, it is hard to subject this locus to algebraic or geometric treatments. We use an affine transformation and map the spectral locus as a tractable geometry.

$$X = \int L_{\varepsilon,\lambda}\, \bar{x}(\lambda) d\lambda$$
$$Y = \int L_{\varepsilon,\lambda}\, \bar{y}(\lambda) d\lambda$$
$$Z = \int L_{\varepsilon,\lambda}\, \bar{y}(\lambda) d\lambda \qquad (2.1a\text{-}e)$$
$$x = \frac{X}{X+Y+Z}$$
$$y = \frac{Y}{X+Y+Z}$$

Define an affine transformation on the CMF's: For the case of monochromatic SRD, let $w_1(\lambda), w_2(\lambda)$, $w_3(\lambda)$ be three auxiliary functions of the wavelength and define the affine tri-stimulus values $(X_a, Y_a, Z_a)$ through Eqn. (2.2a-c)

$$X_a = X + \int L_{\varepsilon,\lambda} w_1(\lambda)\, d\lambda$$
$$Y_a = Y + \int L_{\varepsilon,\lambda} w_2(\lambda)\, d\lambda \qquad (2.2a\text{-}c)$$
$$Z_a = Z + \int L_{\varepsilon,\lambda} w_3(\lambda)\, d\lambda$$

where X, Y and Z are the tri-stimulus values for monochromatic wavelengths (Eqn. (2.1a-c)). Now choose the auxiliary functions so that the spectral locus is mapped as an ellipse in the new system (affine system, $(X_a, Y_a, Z_a)$), centered at a conveniently chosen point (M,N). The algebraic equations defining this transformation are at Eqn. (2.3a-c).

$$\frac{c6\, x(\lambda) + c3\, w1(\lambda)}{c6\, x(\lambda) + c7\, y(\lambda) + c8\, z(\lambda) + c3\, w1(\lambda) + c4\, w2(\lambda) + c5\, w3(\lambda)} = c1 \cos(\theta) + M$$

$$\frac{c7\, y(\lambda) + c4\, w2(\lambda)}{c6\, x(\lambda) + c7\, y(\lambda) + c8\, z(\lambda) + c3\, w1(\lambda) + c4\, w2(\lambda) + c5\, w3(\lambda)} = c2 \sin(\theta) + N \qquad (2.3a\text{-}c)$$

$$c6\, x(\lambda) + c7\, y(\lambda) + c8\, z(\lambda) + c3\, w1(\lambda) + c4\, w2(\lambda) + c5\, w3(\lambda) = P$$

The parameters c1, c2 define the semi-major and semi-minor axes of the target ellipse. Parameters c3, c4, c5 are coefficients of the affine terms $w_1(\lambda), w_2(\lambda)$ and $w_3(\lambda)$. The parameters c6, c7, c8 are the coefficients of the CMF's. These equations perform an affine transformation on $x(\lambda), y(\lambda)$ and $z(\lambda)$ as well as do a normalization of the transformed system with the LHS on Eqn. (2.3c). The normalization is analogous to the defining equations of the chromaticity coordinates (Eqn. (2.1d,e)). LHS of Eqn. (2.3a,b) will be referenced as affine chromaticity coordinates $(x_a, y_a)$.

In the calculations, parameters c3, c4, c5, c6, c7, c8 are set equal to unity so that the shapes of the CMF's and the affine functions are unscaled. Since choosing these c's, the parameters available for choosing appropriate affine functions (w's) are: M, N, P and the target ellipse's axis parameters c1, c2. Now on solving the system (Eqn. (2.3)) for $w_1(\lambda), w_2(\lambda)$ and $w_3(\lambda)$, we obtain Eqn. (2.4a-c). For pure colors, the function CMF+(corresponding w) shall be referred as affine CMF and the w's as affine functions.

$$
\begin{aligned}
w_1(\lambda) &= \frac{c_1 P\cos(\theta) + MP - c_6 x(\lambda)}{c_3} \\
w_2(\lambda) &= \frac{c_2 P\sin(\theta) + NP - c_7 y(\lambda)}{c_4} \\
w_3(\lambda) &= -\frac{c_8 z(\lambda) + c_1 P\cos(\theta) + MP + c_2 P\sin(\theta) + NP - P}{c_5}
\end{aligned}
\qquad (2.4a\text{-}c)
$$

In order to ensure that the affine system is well defined for all choices of monochromatic wavelengths and intensities, it is required that the affine functions are positive valued in the frequency range of interest (visible wavelengths). A parameter combination that satisfies this requirement is at Table 2.1.

Table 2.1 Set Parameters

| Parameter | value |
|---|---|
| M | 0.2 |
| N | 0.1 |
| P | 5.8 |
| c1 | 0.1 |
| c2 | 0.1 |

These functions when added to the color matching functions, the resulting curve has the property that it maps onto the boundary of an ellipse, as required by the Eqn. (2.3). The CMF's for the standard observer (CIE 1964, 10 deg CMF's), pictured in Fig. 2.1 are used for evaluating the functions at Eqn. (2.4). The three affine functions $w_1(\lambda), w_2(\lambda)$ and $w_3(\lambda)$ (Eqns. (2.4a-c)) are plotted in Figs. (2.3-2.5).

The choice M, N determines the center of the ellipse in the affine system, while P governs the intercept of the normalization plane. In general, these parameters may be varied for conveniently positioning the elliptical spectral locus in the affine coordinate system. $\theta$ varies in $[0, \pi]$ and maps bijectively with the wavelength from blue until red. In order to use the full symmetry of the ellipse, the $\theta$ is continued through $2\pi$ and the arc from $[\pi, 2\pi]$ is interpreted as the bijective mapping of the line of purples into the lower half of the circle (marked in cyan, Figs. (2.7), (2.8)).

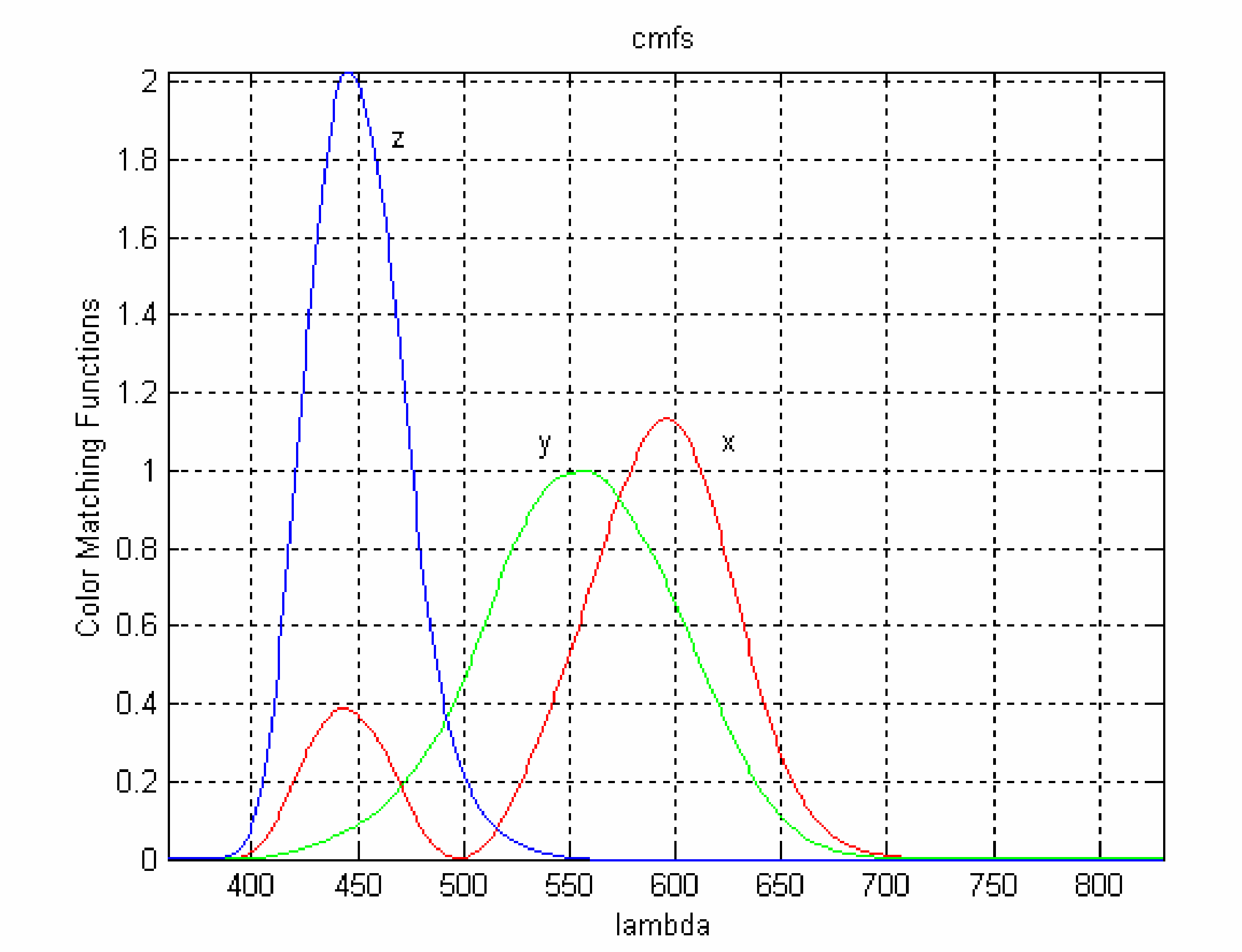

**Fig. 2.1** Color Matching Functions (CIE, standard observer) used for demonstrating the affine transformation. The curves are color coded according to the cones.

Remarks: 1) unlike the CMF's, the affine functions w1, w2, w3 are not normalized. The affine transformation is introduced to remap the (algebraically intractable) spectral locus and the line of purples into a symmetric plane locus and thereby gain symmetry elements (homogenized cones) in the problem. 2) Unlike the CMF's, the affine CMF's may be conveniently approximated using fifth order polynomials 3) The $w_1(\lambda), w_2(\lambda)$ and $w_3(\lambda)$ functions show downward peaks at locations the CMF's exhibit upward peaks. 4) the bijection of $\theta$ with wavelength along the spectral locus is precisely defined by the angular, wavelength and the line (of purples) extents. 5) the affine transformation (Eqns. (2.2), (2.3)) is defined only for pure colors (monochromatic wavelengths).

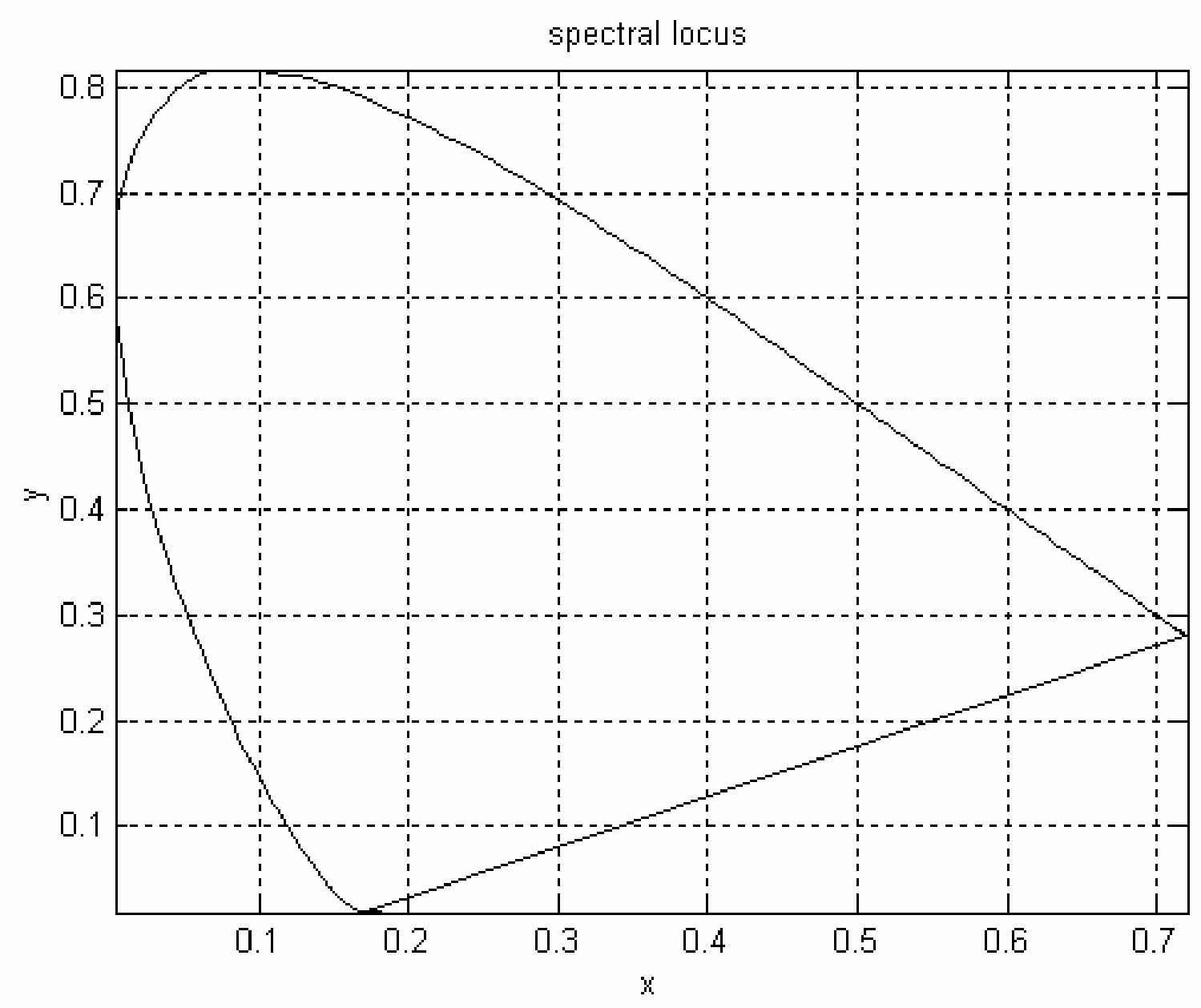

**Fig. 2.2** The boundary of the chromaticity diagram obtained by sweeping a spike (monochromatic SRD) of unit height across the color matching functions and using Eqn. (2.1) for computing the chromaticity coordinates. The straight line is the line of purples and joins the ends of 380 nm (beyond blue) and 760 nm, red. Purples are obtained by combining the red and blue wavelengths in various proportions.

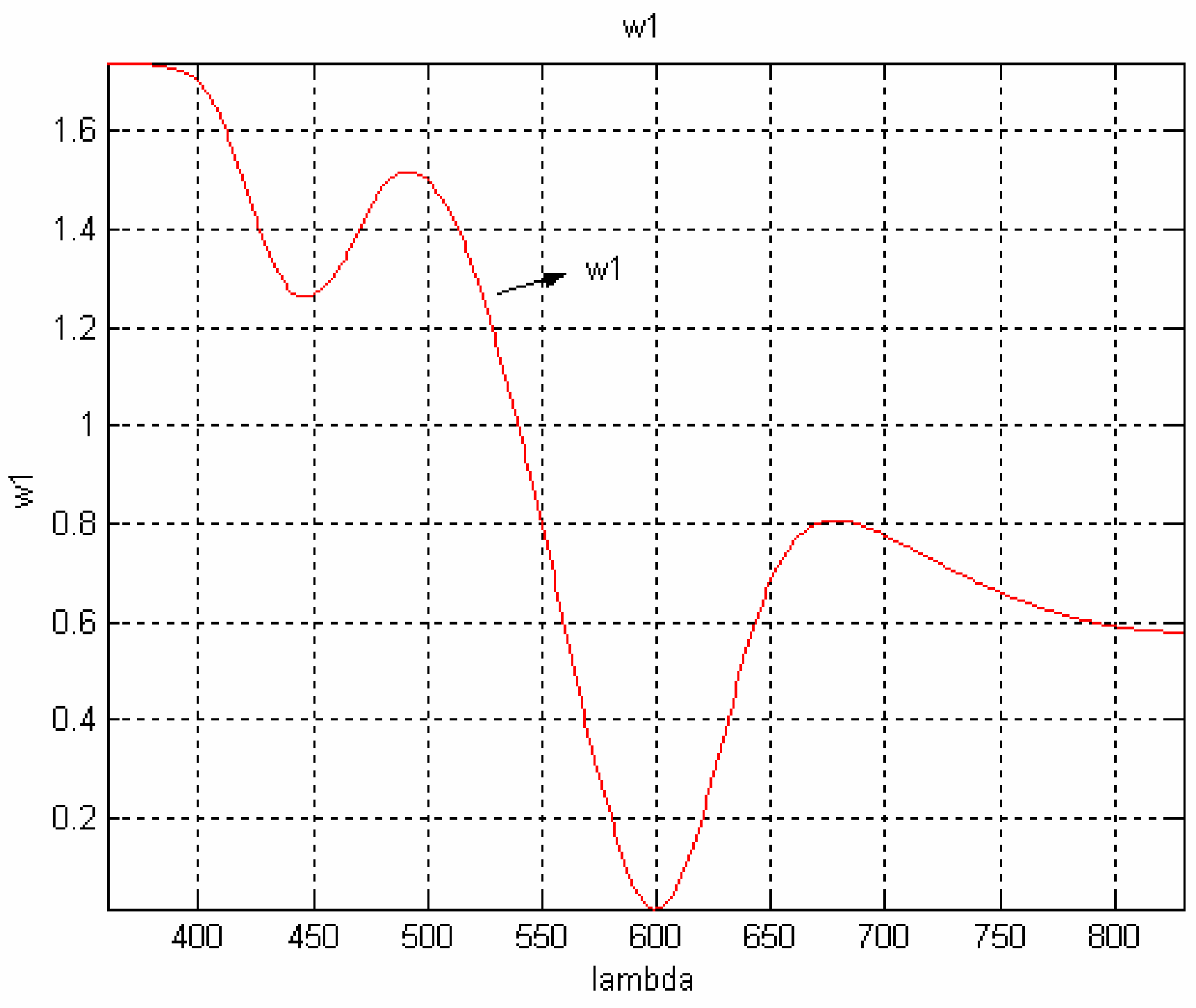

**Fig. 2.3** The affine function ( $w_1(\lambda)$ , Eqn. (2.4a)) for the CMF of the long wavelength cone, drawn in red. At 450nm and at 600 nm, This curve dips corresponding with the peaks of the long cone's CMF.

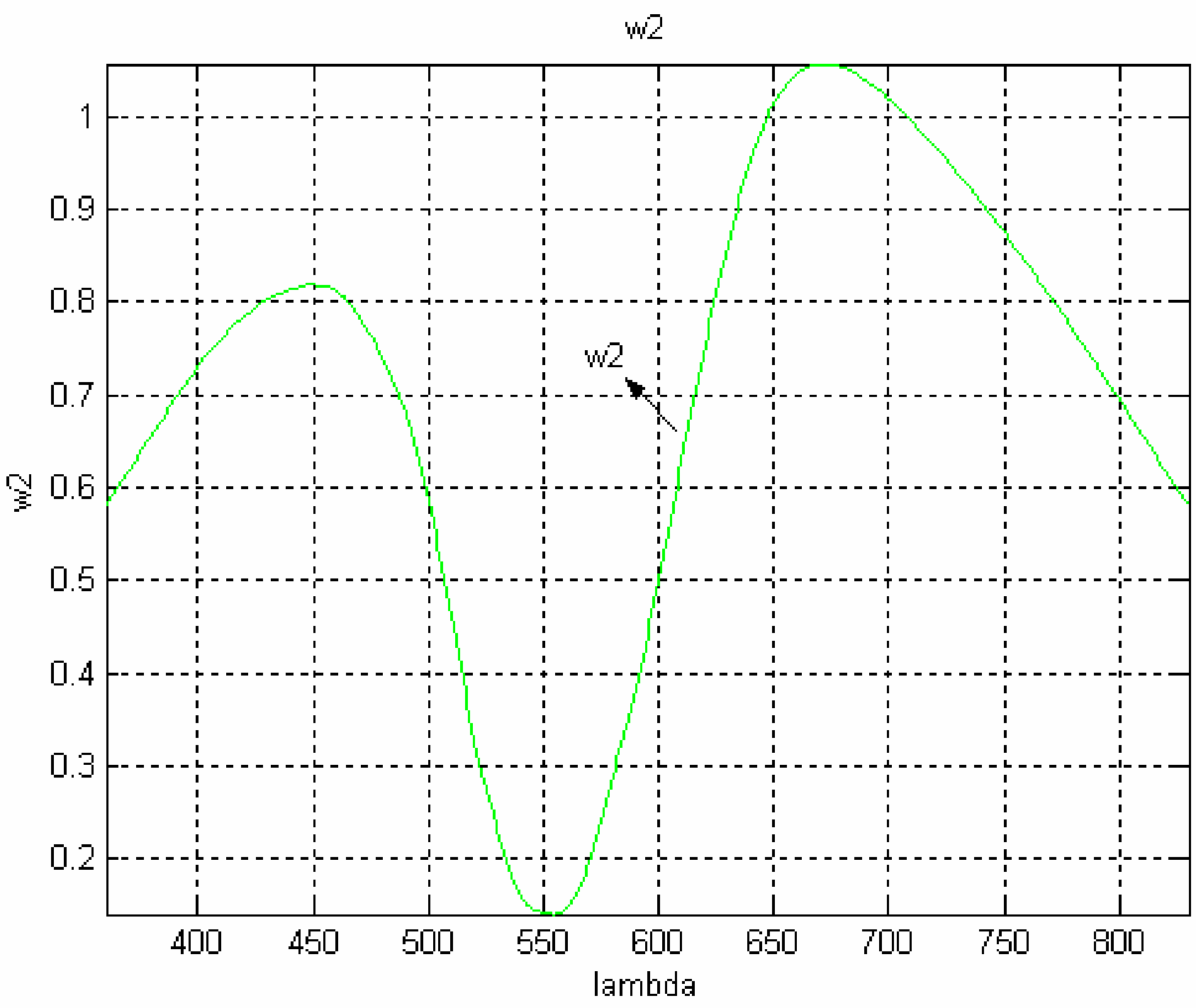

**Fig. 2.4** The affine function ( $w_2(\lambda)$, Eqn. (2.4b)) for the CMF of the middle wavelength cone, drawn in green. The dip at 550nm corresponds to the peak at the same wavelength for the middle cones.

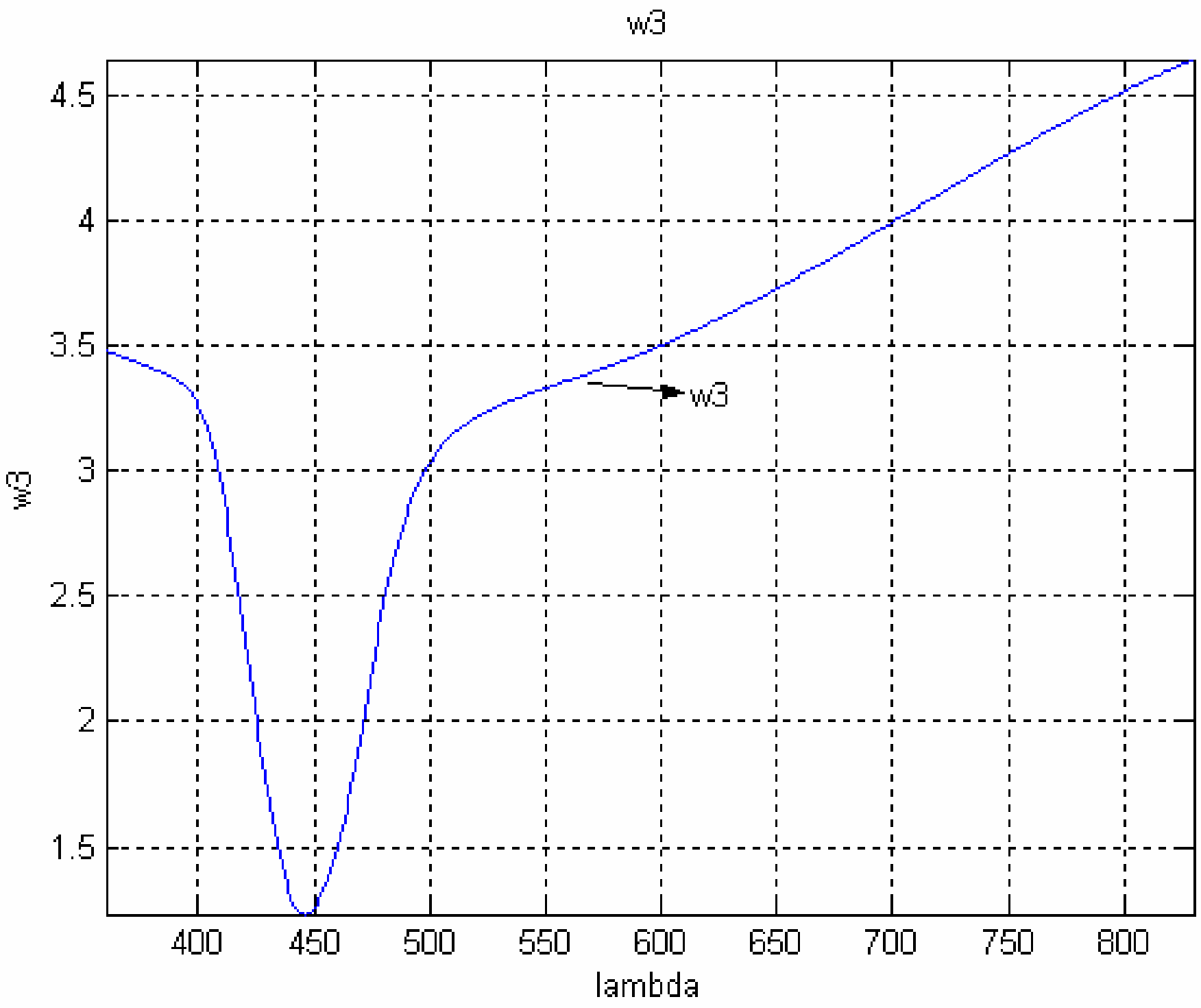

**Fig. 2.5** The affine function ( $w_3(\lambda)$, Eqn. (2.4c)) for the CMF of the short wavelength cone, drawn in blue. The dip at 450 nm corresponds to the peak of the blue cone's CMF.

A plot comparing the profiles of the CMF's and the affine transformed (CMF + corresponding w-function) CMF's is at Fig. 2.6.

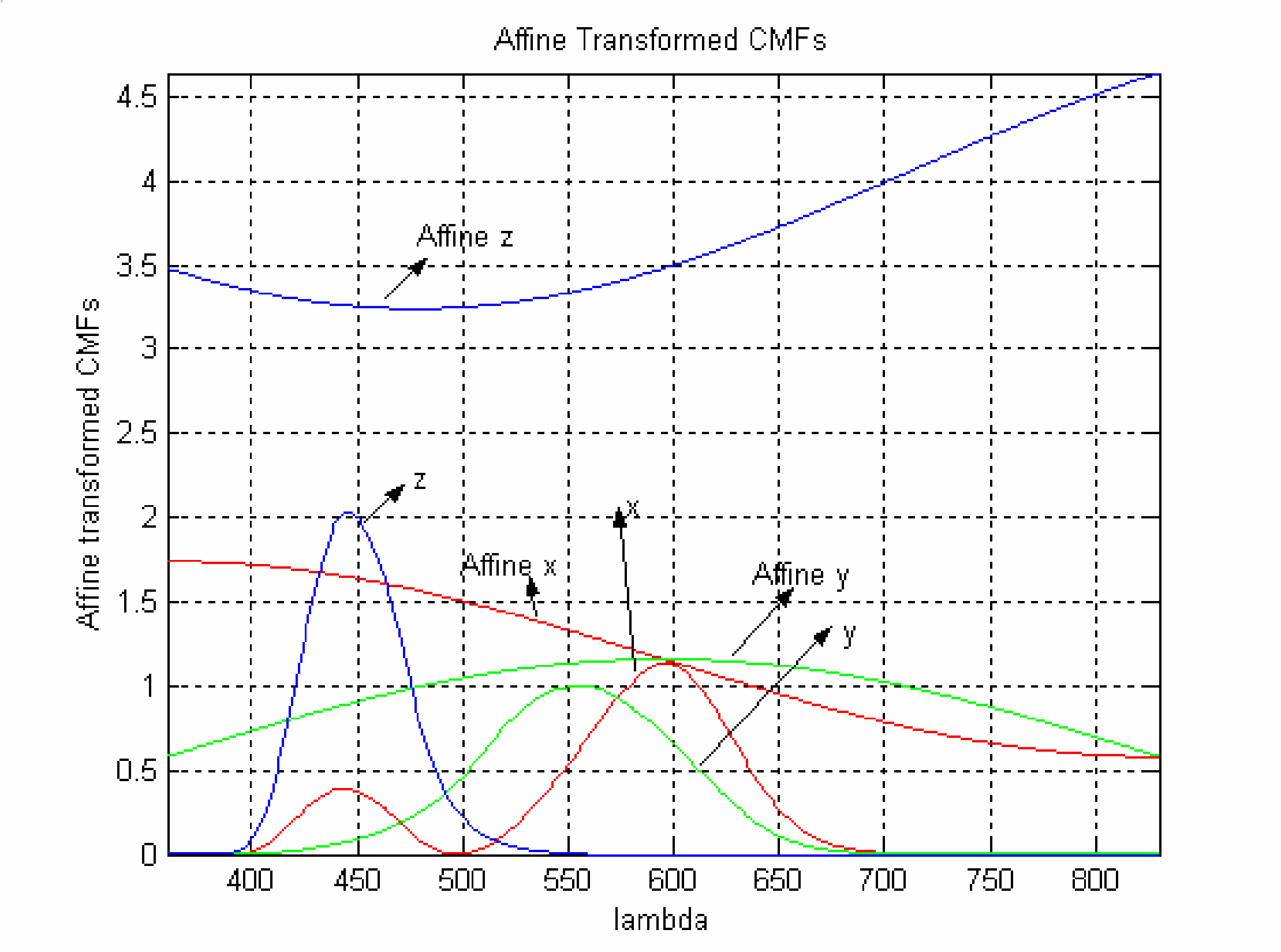

**Fig. 2.6** the affine CMF's are plotted against the wavelength and compared with the CMF's. Plots are color coded. The main purpose of developing the affine functions is to introduce a well-defined (algebraically tractable) geometric shape to the spectral locus. The parameters in the Eqns. (2.3), (2.4) are so set that the spectral locus in the new (affine) system is a circle.

Using the affine transformed CMF's, the construction of the affine spectral locus (ellipse including the arc of purples) from the spectral locus is detailed in Figs. (2.7, 2.8). The angle $\theta$ runs counter clockwise. As marked in the figures, the low wavelengths are mapped near $\theta = 0$ and high wavelengths are mapped near $\theta = \pi$**.** Then upon hitting $\theta = \pi$, the curve is continued until $\theta = 2\pi$, carrying the line of purples as the arc of purples, marked in cyan in Fig. 2.8. Through this construction, the boundary of the chromaticity diagram is mapped as a symmetric and algebraically tractable elliptic locus (circle).

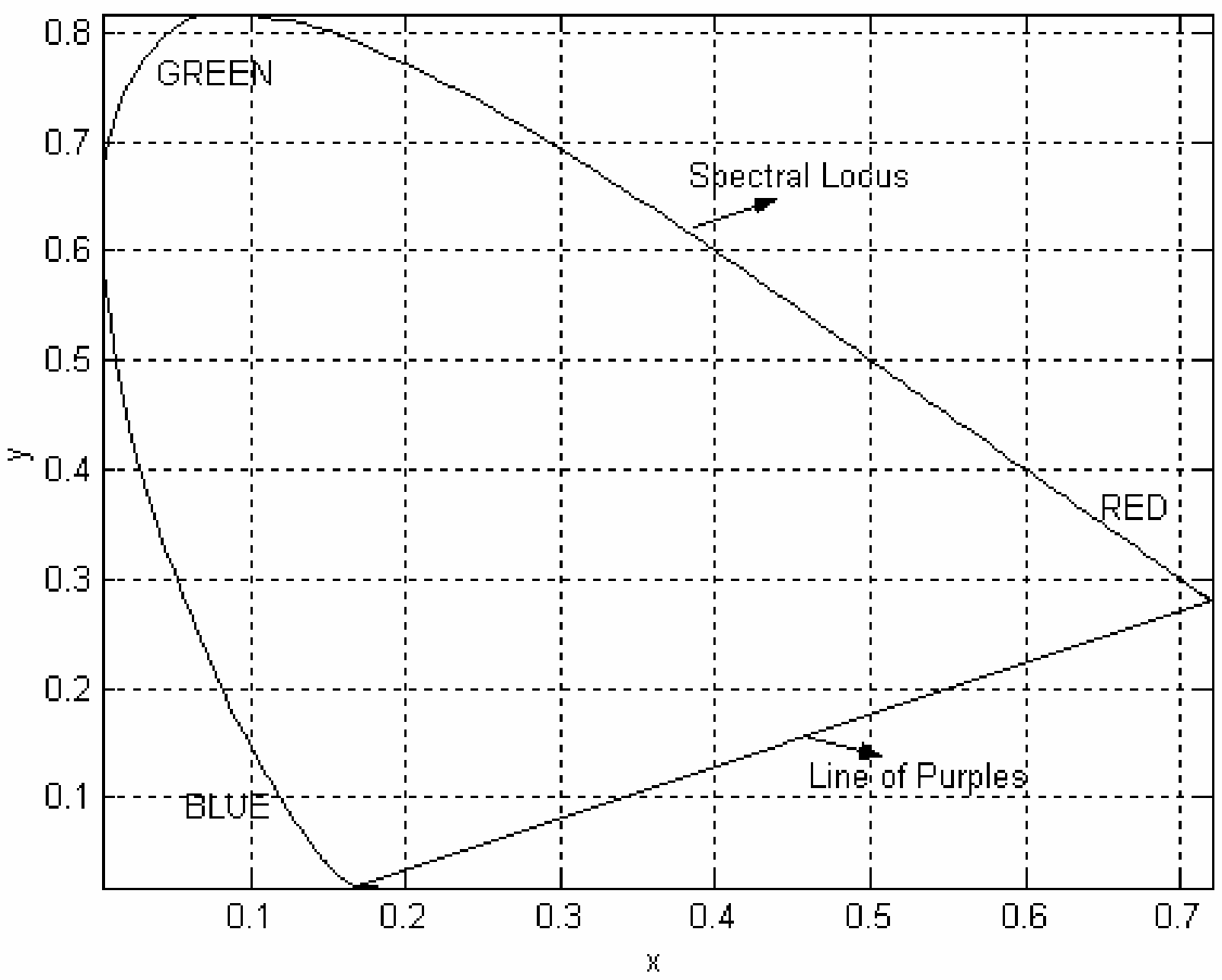

**Fig. 2.7** The spectral locus on the chromaticity diagram and the line of purples make the boundary (for standard observer and within the tolerances and assumptions of the color matching experiment) for color perception. These loci are remapped as a circle using an affine transformation (Fig. 2.8).

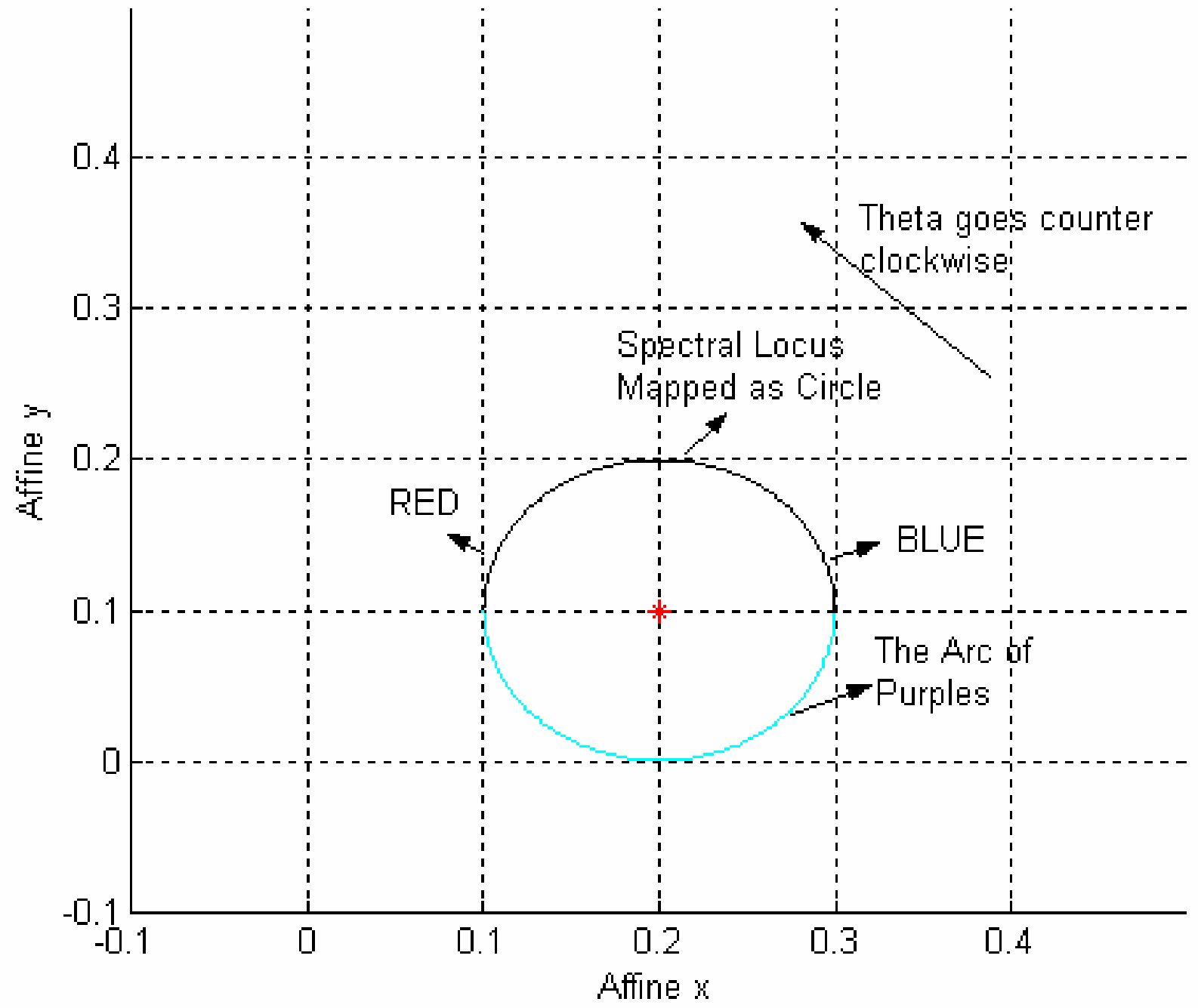

**Fig. 2.8** The angle theta runs counter clockwise. The upper arc (black) is an affine mapping of the spectral locus of the chromaticity diagram as an arc of ellipse (circle). Pure red occurs at theta=$\pi$ and blue occurs near theta=0. The line of purples of the CIE diagram then maps into the circle from theta = [$\pi$,2$\pi$] ("arc of purples", marked in cyan) continues the construction and meets up theta=0. The affine transformation plus the arc of purples together map the boundary of the CIE chromaticity diagram (spectral locus plus line of purples) as a circle in the affine coordinate plane. X-axis is $x_a$, Y-axis is $y_a$.

The analytical equation of the circle is used for homogenizing the erected cylinder with respect to the normalizing plane (Eqn. (2.3c)) and actually constructing the analytical equation to the spectral chromaticity cone in the affine coordinates.

The equation of the circle of Fig. 2.8 is

$$(x_a - M)^2 + (y_a - N)^2 = R^2 \tag{2.5}$$

The circle is centered at $(x_a, y_a)$ =(M,N) on the $x_a y_a$-plane and erects a right circular cylinder into the (Xa,Ya,Za) space, with axis parallel to Za and passing through (M,N,0). The equation of the erected cylinder is (Xa, Ya capitalized, for the surface in (Xa,Ya,Za) space)

$$(X_a - M)^2 + (Y_a - N)^2 = R^2 \tag{2.6}$$

The sectional plane for this cylinder is (from Eqn. (2.3c))

$$X_a + Y_a + Z_a = P = 5.8 \tag{2.7}$$

Upon homogenizing the right circular cylinder (Eqn. (2.6)) with respect to the plane (Eqn. (2.7)), we have the analytical equation of the homogenized cone (Eqn. (2.8), Figs. (2.9), (2.10), (2.11). While homogenizing, we assume that the plane intersects the cylinder. Planes parallel to the cylinder's axis result in imaginary cones.)

$$X_a^2 - \frac{2X_a M(X_a + Y_a + Z_a)}{P} + \frac{M^2(X_a + Y_a + Z_a)^2}{P^2} + Y_a^2 - \frac{2Y_a N(X_a + Y_a + Z_a)}{P} + \frac{N^2(X_a + Y_a + Z_a)^2}{P^2} = \frac{R^2(X_a + Y_a + Z_a)^2}{P^2} \tag{2.8}$$

Substituting M=0.2, N=0.1, R=0.1 and P=5.8 into the equations, we obtain the specific equations of all geometric objects (circle in the plane, cylinder in 3-space, and the homogenized cone in that order, Eqns. (2.9, 2.10, 2.11)) considered for calculations and pictured in the plots starting at Fig. (2.9).

$$(x_a - 0.2)^2 + (y_a - 0.1)^2 = R^2 \tag{2.9}$$

$$(X_a - 0.2)^2 + (Y_a - 0.1)^2 = R^2 \tag{2.10}$$

$$X_a^2 - 0.06896551724X_a(X_a + Y_a + Z_a) + 0.001486325802(X_a + Y_a + Z_a)^2 + Y_a^2 - 0.03448275862Y_a(X_a + Y_a + Z_a) = 0.0002972651605(X_a + Y_a + Z_a)^2 \tag{2.11}$$

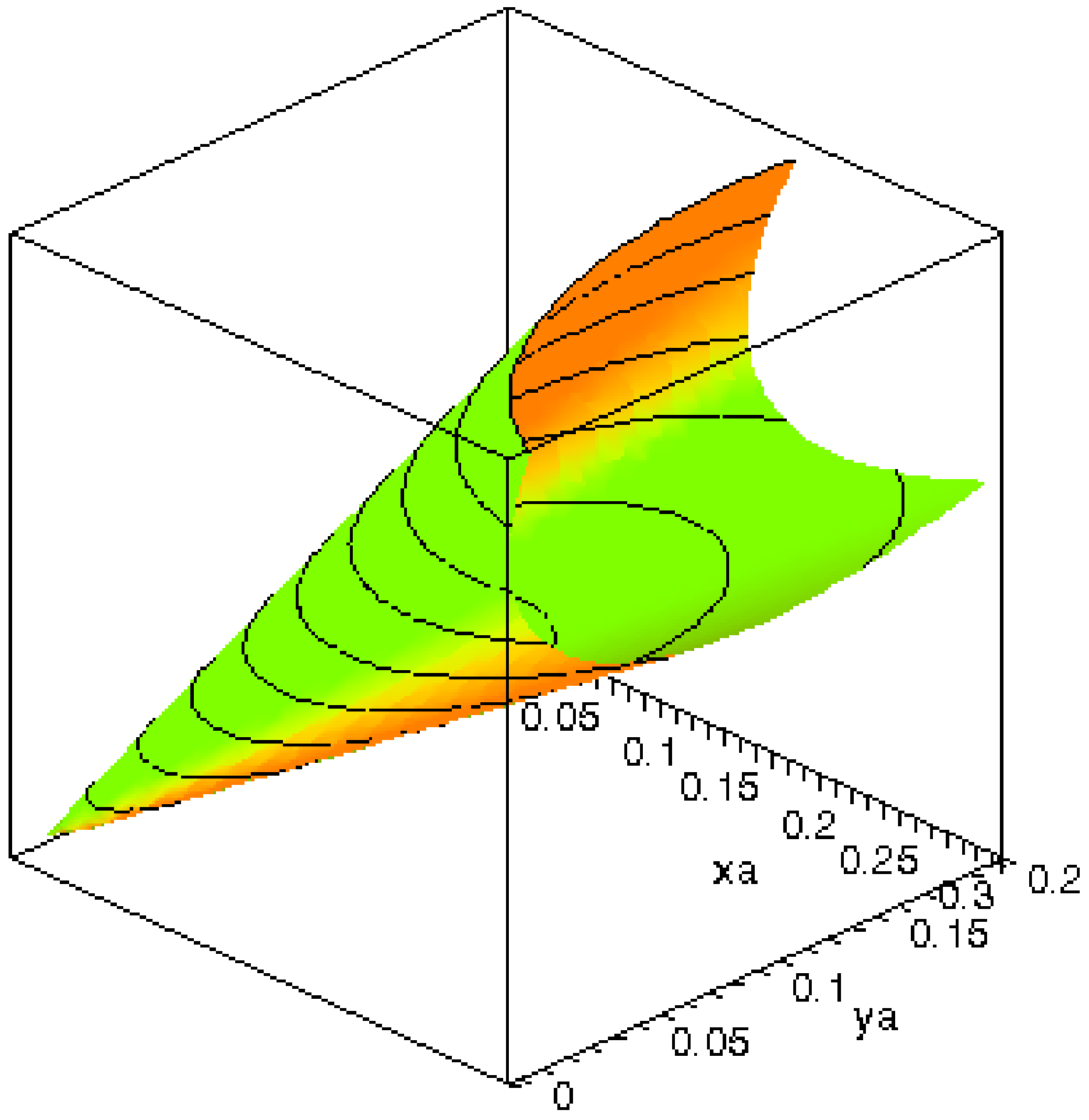

**Fig. 2.9** The spectral chromaticity cone obtained by homogenizing the cylinder (Eqns. (2.6), (2.10)) with respect to the normalizing plane Xa+Ya+Za=5.8 (Eqn. (2.3c)). Contour plot is shown here. (Eqn. (2.11) is its analytical description.)

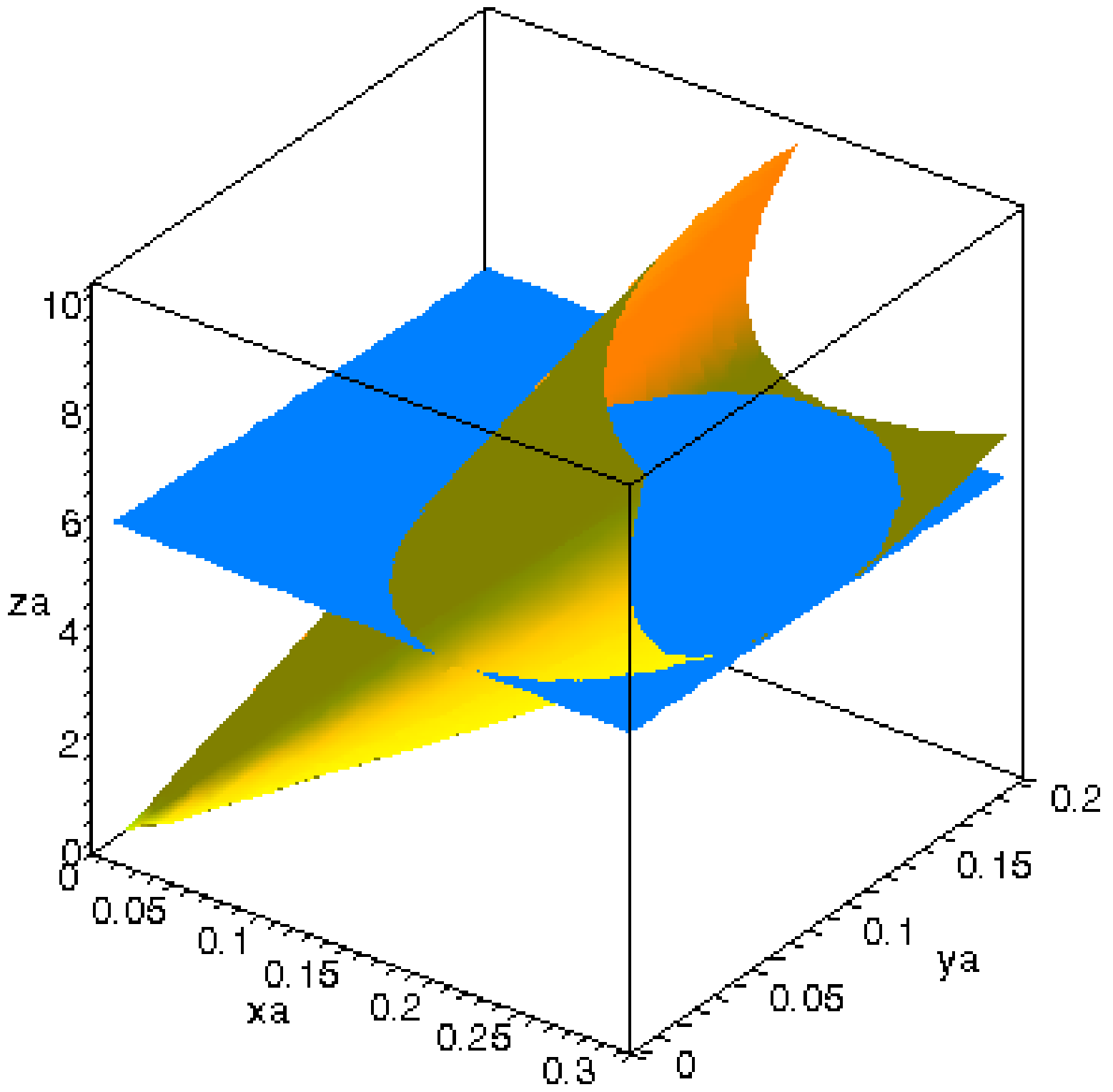

**Fig. 2.10** The section of the homogenized cone (gold) with the plane Xa+Ya+Za=5.8 (cyan) is a circle in the top view (judging from the boundary of cyan over yellow, in the cone's interior). Analytical equation of the cone is at Eqn. (2.11)

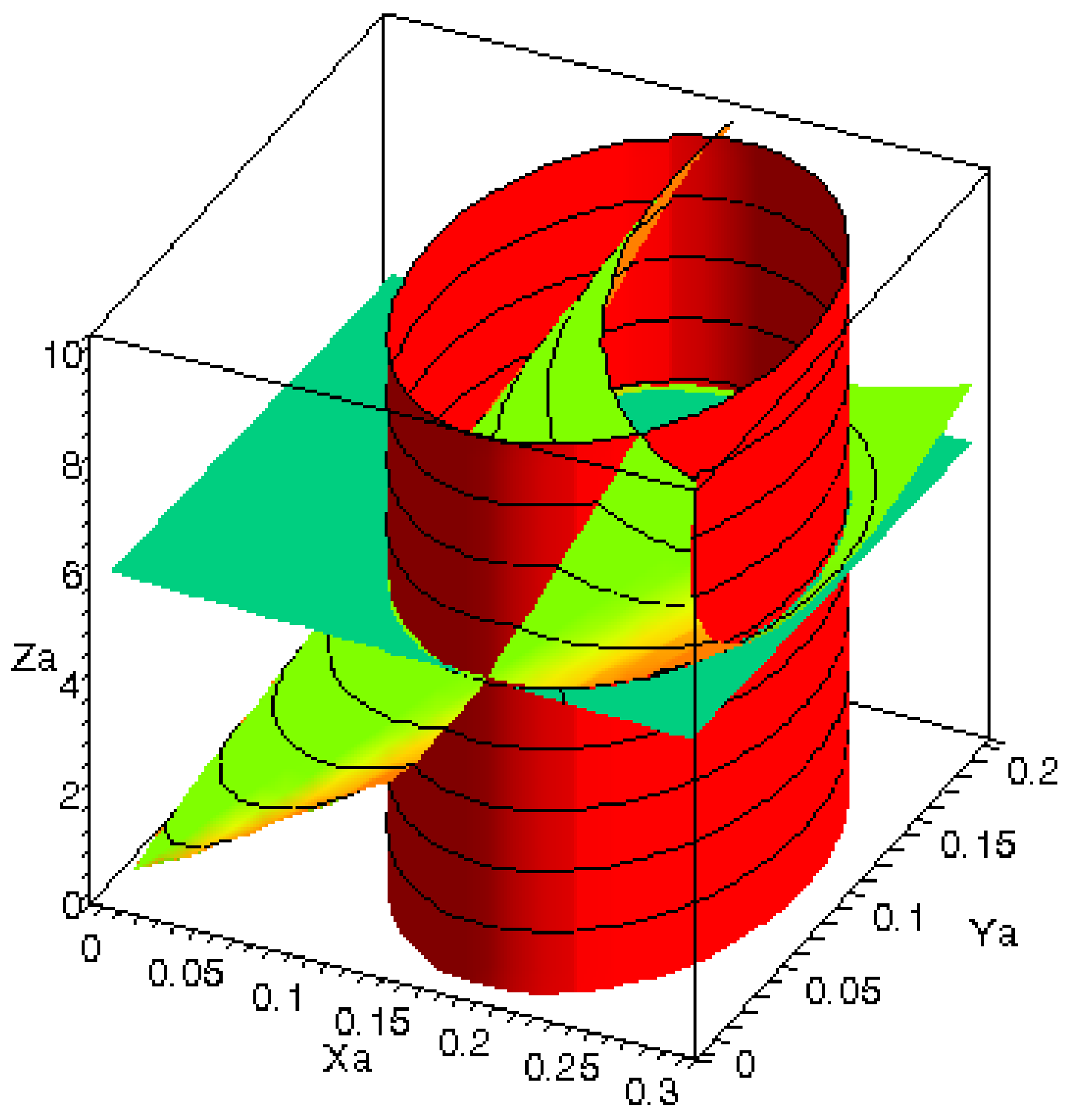

**Fig. 2.11a** The geometric situation of homogenizing a cylinder (red) with respect to a plane is pictured. The resulting homogenized surface is a cone (green) with apex at the origin and passing through the intersection of the plane with the cylinder. In the present example, the axis line of the cylinder passes through (0.2,0.1,0) parallel to Za-axis. Radius of the cylinder is 0.1. The plane has equal intercepts of 5.8 on all three axes. This is the geometric situation of the spectral chromaticity cone in the affine coordinate system obtained by adding affine terms, w1, w2, w3 to the standard color matching functions, according to Eqn. (2.3a-c).

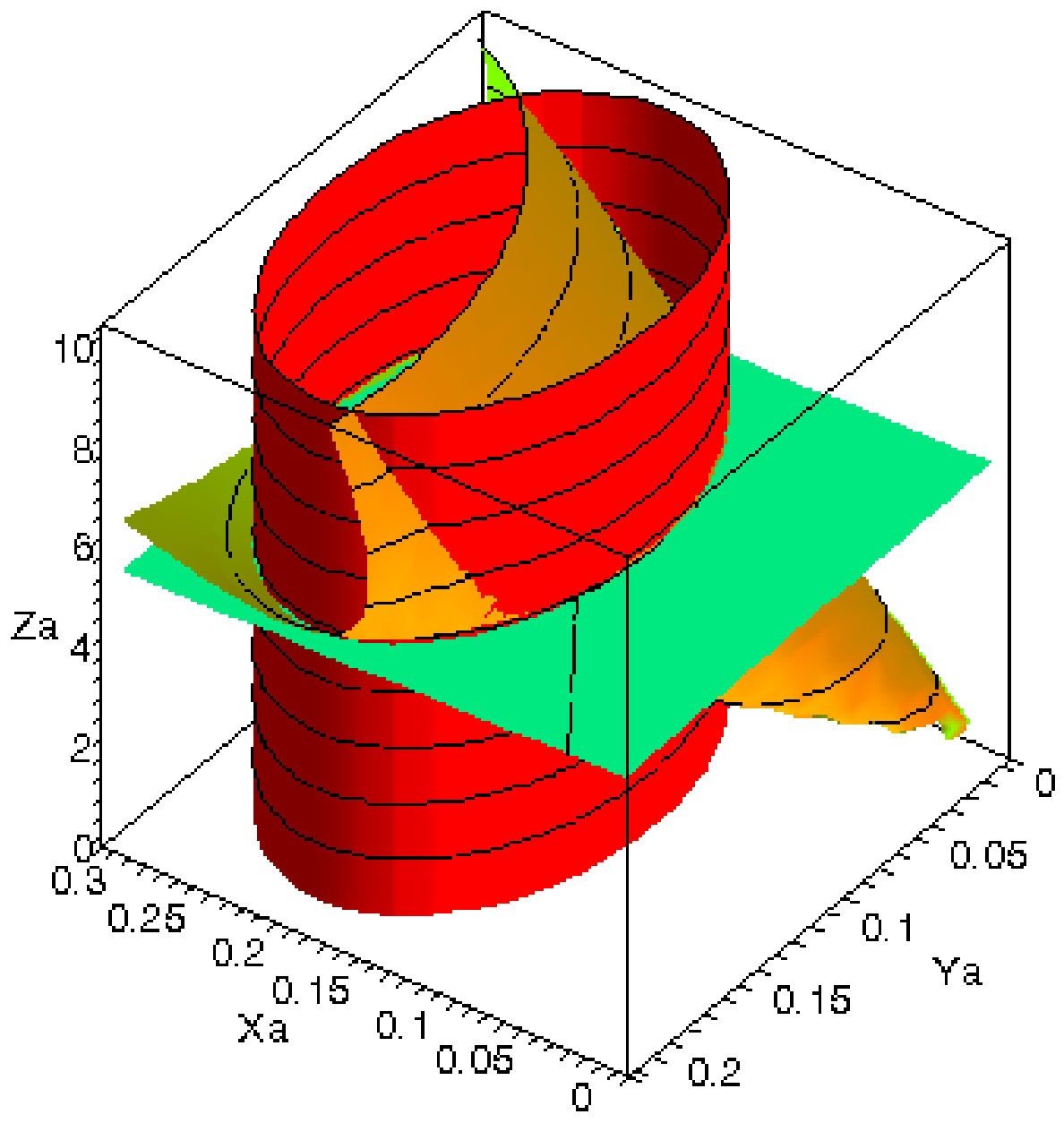

**Fig.2.11b** View 2 of the homogenized cone.

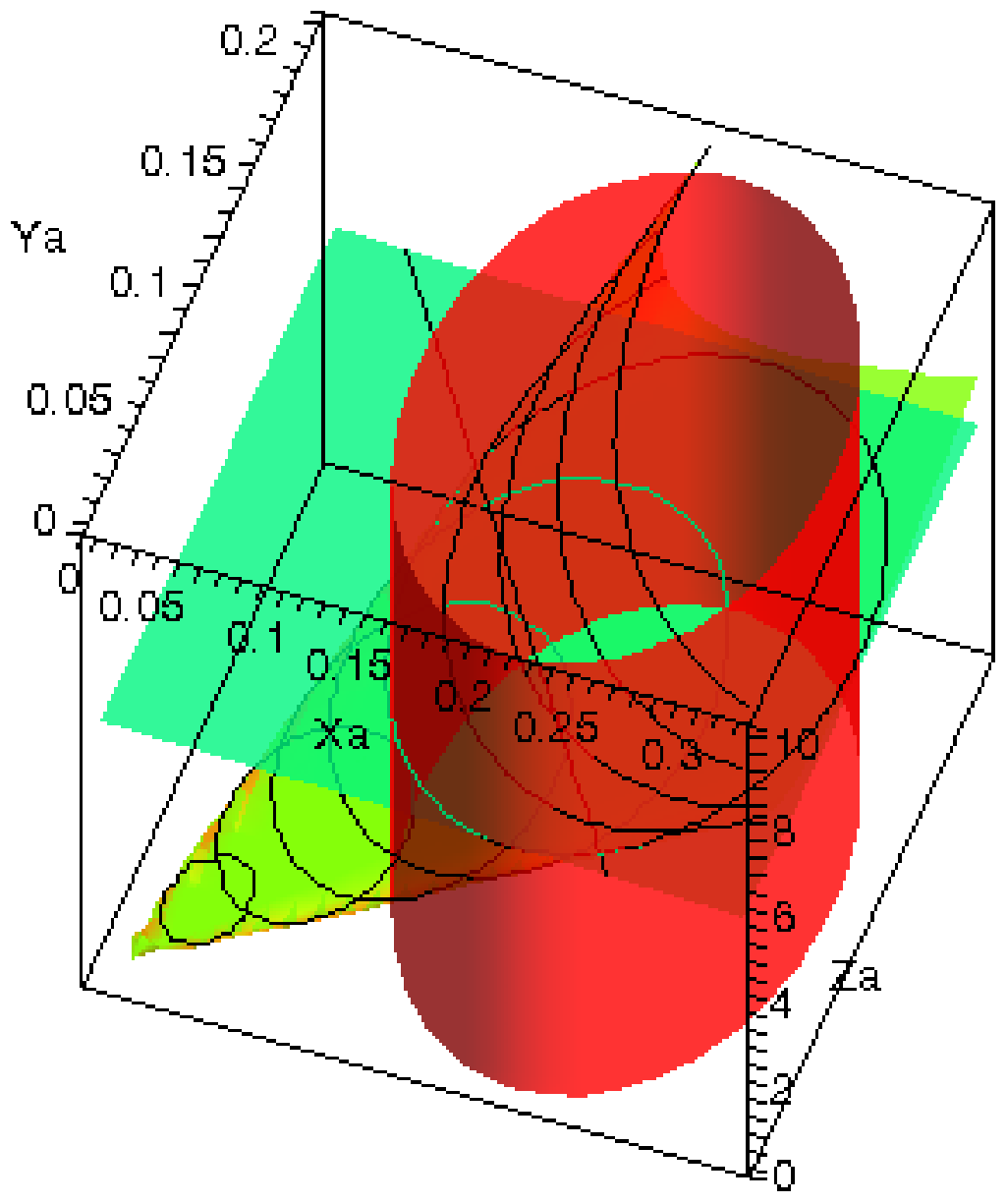

**Fig.2.11c** View 3 of the homogenized cone. Increased transparency shows the details of the intersecting surfaces.

As the final step in the construction of the spectral chromaticity cone, we use the rotation matrix formula for arbitrary axis rotation in space ([3]) for aligning the symmetry axis (the central axis) of the spectral chromaticity cone (of Fig. (2.9)) along the Zas axis (subscript s for symmetry). We obtain a rotation matrix from (Xa,Ya,Za) frame into (Xas,Yas,Zas) frame so that the Zas is the axis of the cone. If the direction cosines of the arbitrary axis in space are (l,m,n) and $\alpha$ is the required angle of rotation then with the convention of the right hand screw rule, the rotation matrix Q for the rotation is ([3], Eqn. (2.12))

$$Q = \begin{bmatrix} l^2(1-\cos(\alpha))+\cos(\alpha) & lm(1-\cos(\alpha))-n\sin(\alpha) & \ln(1-\cos(\alpha))+n\sin(\alpha) \\ lm(1-\cos(\alpha))+n\sin(\alpha) & m^2(1-\cos(\alpha))+\cos(\alpha) & mn(1-\cos(\alpha))-l\sin(\alpha) \\ \ln(1-\cos(\alpha))-n\sin(\alpha) & mn(1-\cos(\alpha))+l\sin(\alpha) & n^2(1-\cos(\alpha))+\cos(\alpha) \end{bmatrix} \tag{2.12}$$

From Fig.(2.8) using the right hand rule, the direction cosines of the required line are l=-sin(beta), m=-cos(beta), n=0. where $\beta = \tan^{-1}(\frac{0.1}{0.2}) = 0.4636 = 26.56 \ \deg$. The required axis points in the fourth quadrant. Therefore l=-0.4472, m=-0.8944, n=0. The angle $\alpha$ that the axis of the cone must rotate (to go in the matrix at Eqn. (2.12)) to align with the vertical symmetry axis is computed from the point of intersection of the cylinder's axis (Fig. (2.11)) with the plane Xa+Ya+Za=5.8. Projecting (0.2,0.1,0) on the plane, we compute Za=5.8-0.2-0.1=5.5. We note from the geometry that $90-\alpha = \tan^{-1}(\frac{5.5}{\sqrt{0.2^2+0.1^2}}) = 1.53 = 87.67\deg$ and therefore $\alpha = 2.3281\deg$. The rotation matrix Eqn. (2.12), upon plugging these values is

$$Q = \begin{bmatrix} 0.9993385861 & 0.0003306818 & -0.03636179977 \\ 0.0003306818 & 0.9998346088 & 0.01818089988 \\ 0.03636179977 & -0.01818089988 & 0.9991732452 \end{bmatrix} \tag{2.13}$$

The transformation

$$\begin{pmatrix} Xa \\ Ya \\ Za \end{pmatrix} = Q \begin{pmatrix} Xas \\ Yas \\ Zas \end{pmatrix} \tag{2.14}$$

maps the (Xa,Ya,Za) frame into the symmetric (Xas,Yas,Zas) frame (consequently, the cone's axis of symmetry coincides with the Zas axis of this frame).

## 3. Homogenizing Cones for the Subsets of the Chromaticity Cone

To capture the interior of the chromaticity cone which is determined by SRD's of all shapes excluding the monochromatic type, we may use the steps described in the previous section, homogenize the three dimensional cylinders erected over plane algebraic curves (algebraic varieties of dimension one) in the chromaticity diagram's interior, with respect to suitable sectional planes and develop homogenized cones from the intersection of the two surfaces. Since the construction steps are clear, we demonstrate an important application of the interior chromaticity cones: the mechanism through which Macadam ellipses occur on the CIE diagram.

We begin with right circular cylinders in XYZ space with axis line parallel to the Y-axis. The section of these cylinders with Y=Y1 (Y1 is the constant luminance) defines the Macadam circle on the constant luminance planes. Upon homogenizing the cylinder with respect to the plane Y=Y1 we obtain the analytical equation to the homogenized cone passing through the circle of intersection (Fig. 3.1), and apex at the origin. This cone is then sectioned with the normalizing plane X+Y+Z=1. The analytical equation to the Macadam ellipse is obtained by eliminating Z from X+Y+Z=1 and the homogenized cone. These steps are algebraically described from Eqns. (3.1-3.6).

Analytical equation of the right circular cylinder of radius R, perpendicular to the constant luminance plane Y=Y1 and whose axis of symmetry passes through (X1,Y1,Z1) (parallel to Y-axis) is given by

$$(X - X_1)^2 + (Z - Z_1)^2 = R^2 \tag{3.1}$$

On expanding we get

$$X^2 - 2XX_1 + X_1^2 + Z^2 - 2ZZ_1 + Z_1^2 = R^2 \tag{3.2}$$

Homogenizing Eqn. (3.2) with respect to the constant luminance plane Y=Y1, we have the homogenized cone

$$X^2 - \frac{2XX_1Y}{Y_1} + \frac{X_1^2Y^2}{Y_1^2} + Z^2 - \frac{2ZZ_1Y}{Y_1} + \frac{Z_1^2Y^2}{Y_1^2} = \frac{R^2Y^2}{Y_1^2} \tag{3.3}$$

Now on eliminating the Z variable from Eqn. (3.3) and Eqn. (3.4),

$$X + Y + Z = 1 \tag{3.4}$$

the eliminant is

$$\begin{aligned} &2X^2Y_1^2 - 2XX_1YY_1 + X_1^2Y + 2Y_1^2XY - 2Y_1^2X + Y_1^2Y + Y_1^2 + 2Z_1YY_1X + 2Z_1Y^2Y_1 - 2Z_1Y^2Y_1 \\ &+ Z_1^2Y - R^2Y^2 = 0 \end{aligned} \tag{3.5}$$

This is a general quadratic equation in two variables. Using the standard methods of algebra, this may be easily shown to be an ellipse (by adding and subtracting terms, regrouping and completing the squares. Here, from the geometric situation we're certain that the conic section is an ellipse.) We plug in specific numbers into these equations and express this situation as intersecting surfaces. We choose R=1, X1= 2, Y1= 2.5, Z1= 3. The geometric situation of the Macadam ellipse is pictured in Fig. (3.1a-c).

The right circular cylinder (red) intersects the constant luminance plane Y=Y1 (blue) in a circle (Fig. (3.1)). By homogenizing the cylinder with respect to the constant luminance plane, we get the homogenized cone through the circle of intersection and apex at the origin (green). The homogenized cone is sectioned by the normalizing plane X+Y+Z=1 (cyan) and the section is a Macadam ellipse. For the choices made, the ellipse obtained is pictured in Fig. (3.2).

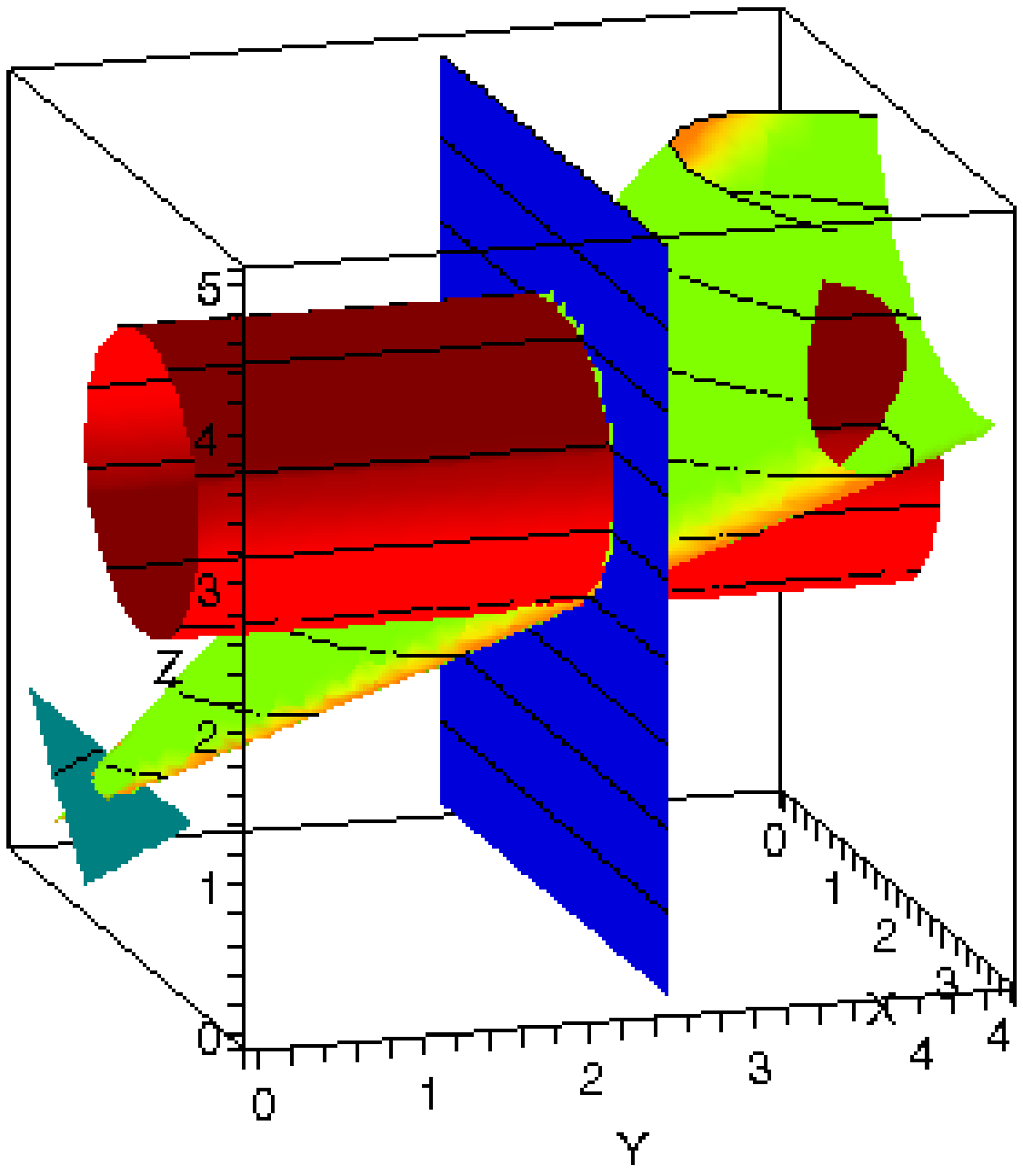

**Fig. 3.1a.** The right circular cylinder intersects (in red. Macadam neighborhood on the constant luminance plane is a circle) the constant luminance plane (blue) and upon homogenizing the cylinder with respect to the plane, we get the homogenized cone (green) whose apex is at the origin and passes through the intersection (circular) of the cylinder and the constant luminance plane. The section of this cone with the normalizing plane (X+Y+Z=1, cyan) is the Macadam ellipse (Fig. (3.2)).

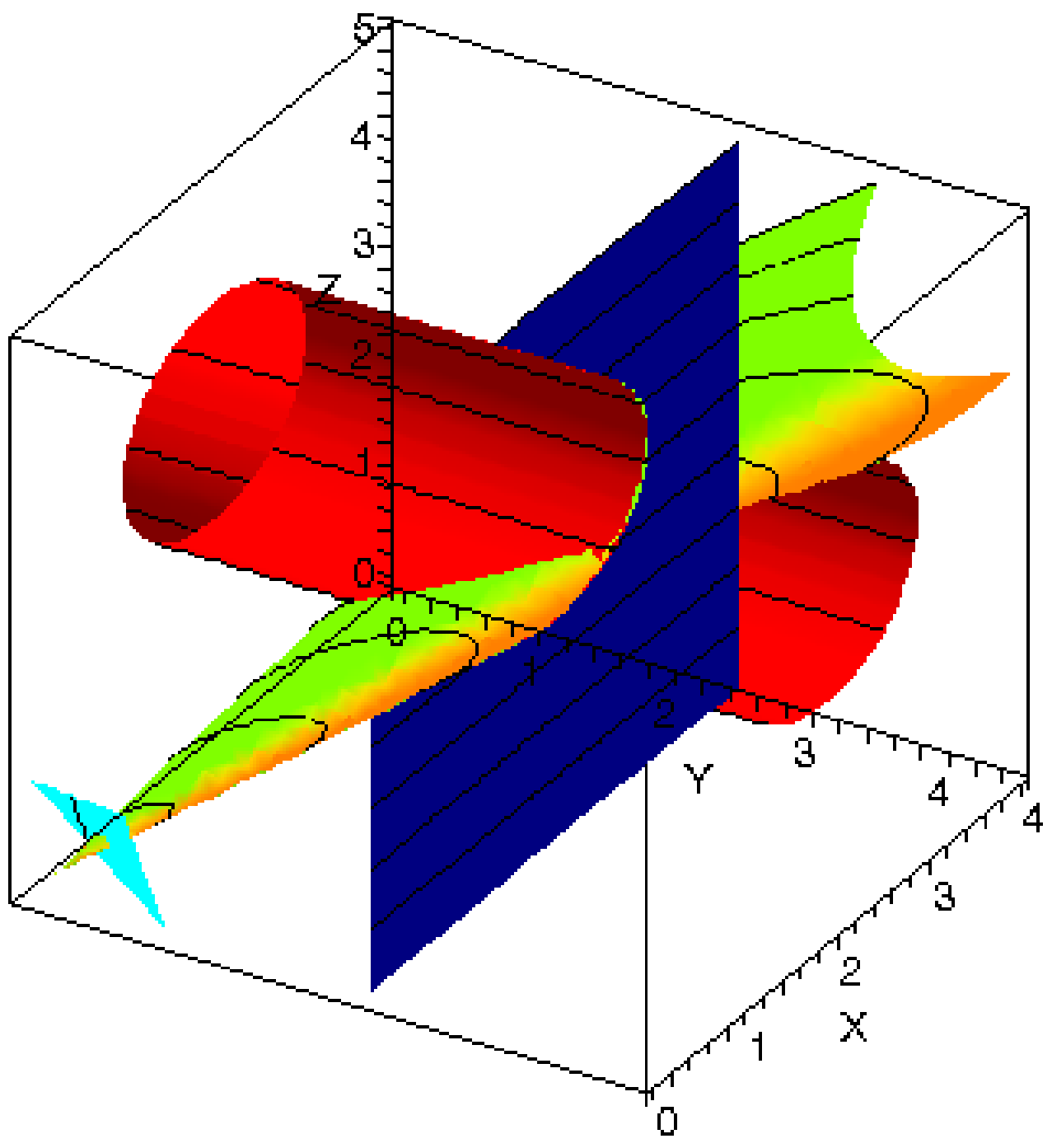

**Fig. 3.1b** View 2 of the homogenizing cone (green)

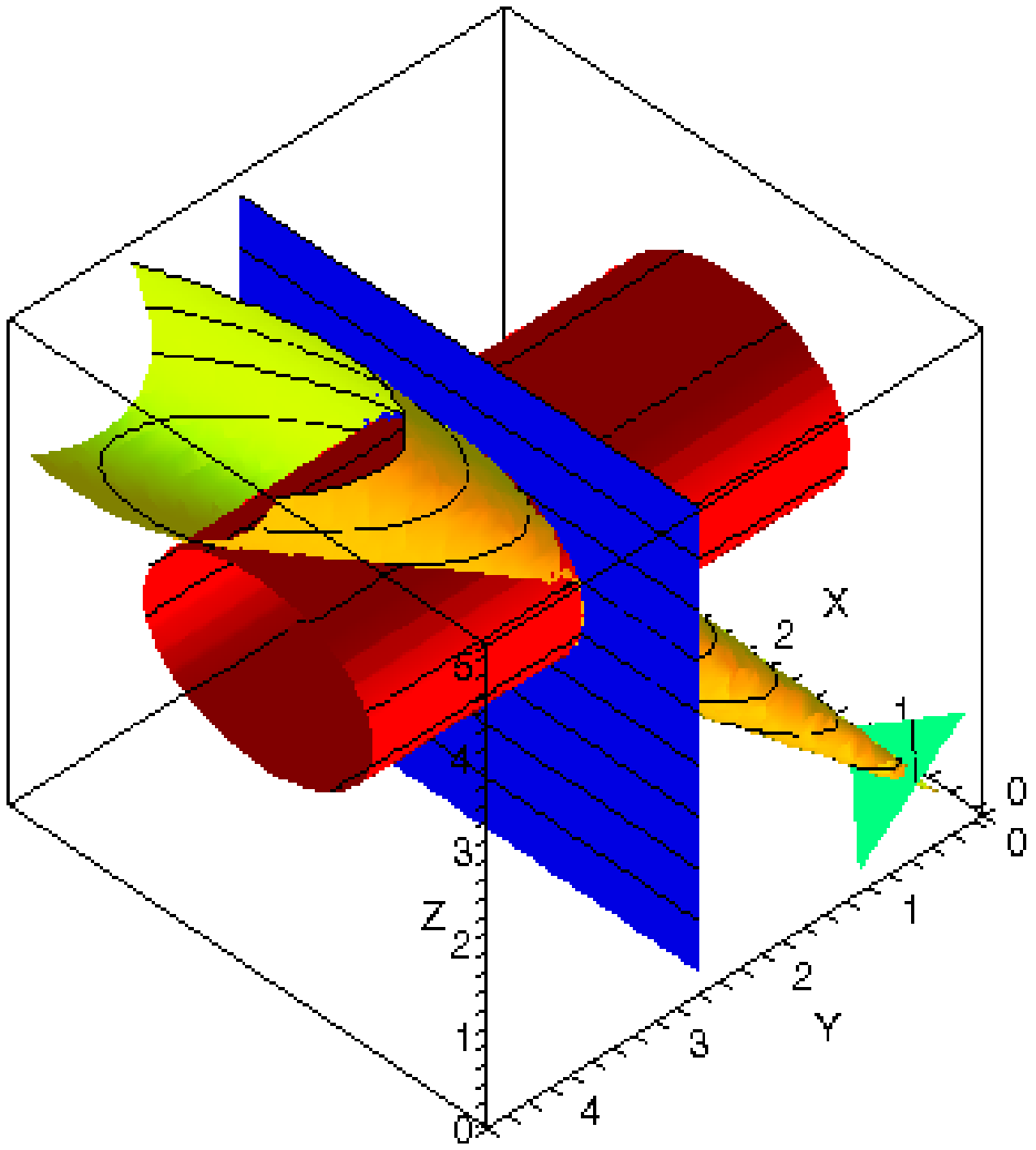

**Fig. 3.1c** View 3 of the homogenizing cone (green, gold lighting)

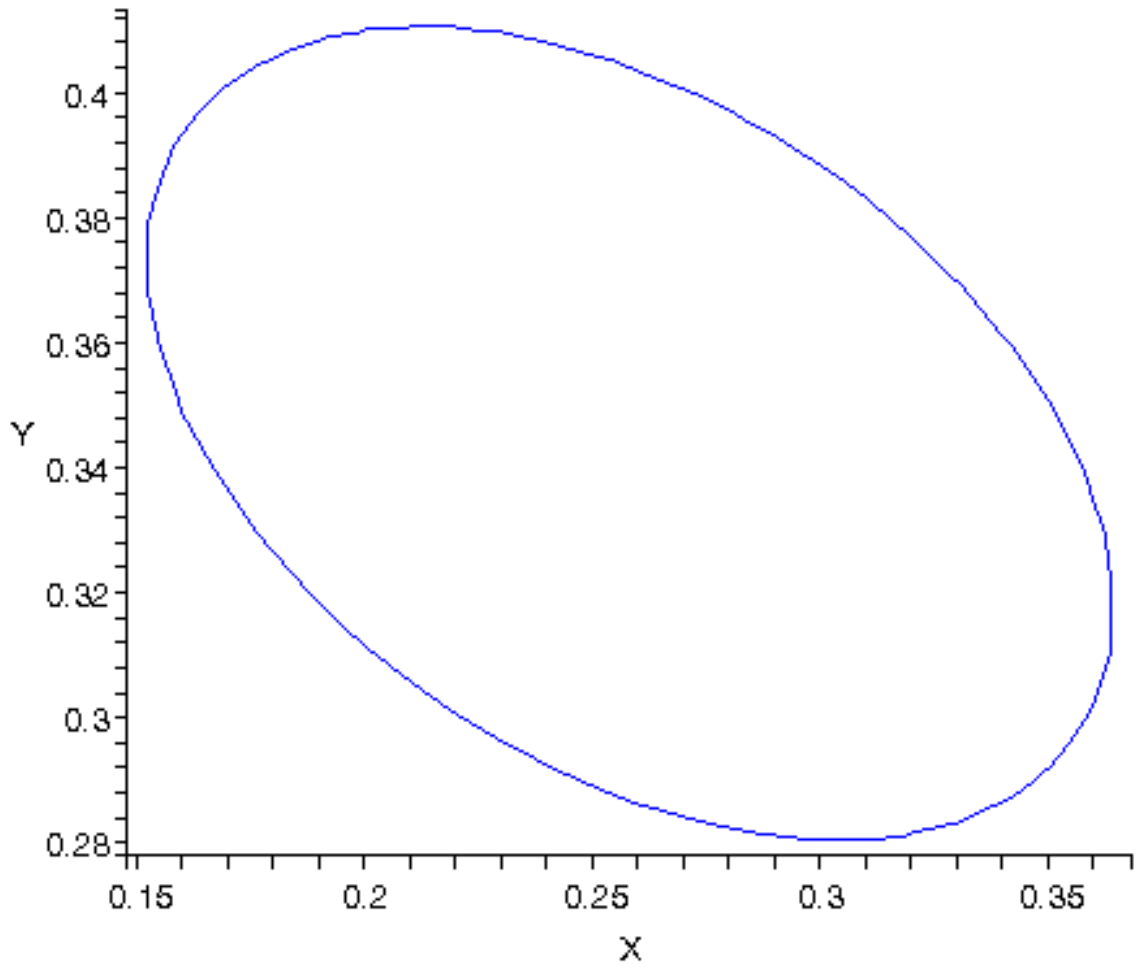

**Fig. 3.2** The Eqn. (3.6) is the intersection of X+Y+Z=1 and the homogenized cone. The plot is an ellipse. Macadam ellipses occur this way.

$$12.50X^2 + 17.50XY + 33.25Y^2 + 6.25 - 12.50X - 27.50Y = 0 \qquad (3.6)$$

Analytical equation of the Macadam ellipse is at Eqn. (3.6). We conclude that the Macadam ellipses are the plane sections (sectioned by X+Y+Z=1) of homogenized cones obtained by homogenizing right circular cylinders with respect to the constant luminance planes. Since this construction is in the interior of the chromaticity cone, the description is entirely in the XYZ notation.

## 4. New Color Order Systems

The transformations described in the previous sections express the spectral chromaticity cone in the simplest form by aligning the z-coordinate axis along the cone's symmetry axis. In the following, we use the cone equation for proposing new types of color orders (the right circular cone of semi-vertical angle 45deg):

$$X_{as}^2 + Y_{as}^2 = Z_{as}^2 \qquad (4.1)$$

a) Helical Ordering of Spectral Colors

The equation of the cone (Eqn. (4.1)) may be used for defining a helical ordering of pure colors. This has the advantage that the light intensity and the (monochromatic) wavelengths of the visible light may be coupled through a differential geometric object and expressed as a color order. This proposal offers a generalization and abstraction for the concept of color ordering. Following is a vector equation for the helix on the cone (Eqn. (4.1), obtained as the intersection of the spiral-cylinder with the cone)

$$R(t) = t\cos(t)I + t\sin(t)J + tK \quad t1 \le t \le t2 \tag{4.2}$$

with I, J and K the unit vectors along the Xas, Yas and Zas axes respectively. The choices t1, t2 define the extent of the helix on the cone.

b) Ordering Colors using Chaotic Dynamical Systems

By defining chaotic dynamical systems on the cone (Eqn. (4.1), example: any discrete-time chaotic dynamical system of the annulus symmetric about the Zas-axis and project on the cone), pure colors may be ordered according to the evolution of the dynamical system and coded by the initial condition of the dynamical system. For non-spectral and unsaturated colors, homogenized cones out of suitably chosen elliptic cylinders and sectional planes may be used. A suitable construction (using annular regions about the symmetry axes of those cones) may define a chaotic dynamical system on the interior cones. For this case, a color order should be viewed as the evolution of "equivalence classes" of SRD's because each tri-stimulus triplet represents a large class of SRD's.

c) Ordering Colors through Stochastic Processes

Any stochastic process defined by the Ito stochastic differential equation ([4]) on the annulus (symmetric about the Zas-axis) and projected on the cone defines a stochastic ordering for pure colors. This ordering is characterized (and coded) by the density function of the stochastic process.

## 5. Conclusion

In the interior of the (CIE) chromaticity diagram, by homogenizing cylinders (erected by algebraic curves) with respect to suitably chosen (sectional or normalizing or both) planes, analytical equations to subsets of the chromaticity cone are developed. The mechanism of occurrence of Macadam ellipses is detailed. For the case of spectral (pure) colors, an affine transformation is used to map the boundary of the (CIE) chromaticity diagram as a circle. By homogenizing the cylinder erected by the circle in three dimensions, with respect to the sectioning (also normalizing) plane, the analytical equation of the spectral chromaticity cone is developed. Using the newly developed equations of the cones, new types of color orders are proposed.

## 6. Acknowledgment

At the RIKEN Brain Science Institute (Saitama, Japan), author likes to thank Cees van Leeuwen. Thanks to MAPLE 11 for the excellent three dimensional plots.

## References

[1] http://www.cvrl.org/
[2] Kaiser, KP and Boynton, MB., (1996), "Human Color Vision," second edition, Optical Society of America, Washington DC.
[3] Craig, JJ, (2004), "Introduction to Robotics: Mechanics and Control," third edition, Prentice Hall.
[4] Oksendal, B., (2003), "Stochastic Differential Equations," sixth edition, Springer.

## Appendix

```
% MATLAB code for Eqns (2.3), (2.4)
% Author has patented the method discussed in this article
load c:/prash/ndCMF.txt; % CMF's available at reference [1]
mat = ndCMF;
lambda = mat(:,1);
coneLong = mat(:,2);
coneMid = mat(:,3);
coneShort = mat(:,4);
lambdaMax = max(lambda);
theta = linspace(0,pi,size(lambda,1))';
c1 = 0.1; % make sure that c1^2+c2^2 <1
c2 = 0.1; c3=1; c4=1; c5=1; c6 = 1; c7 = 1; c8=1;
C=5.8;
A=0.2; B=0.1;
x = coneLong;
y = coneMid;
z = coneShort;

% the affine curves...
w1 = (c1 * cos(theta) * C + C*A- c6 * x) / c3;
w2 = (c2 * sin(theta) * C + C*B - c7 * y) / c4;
w3 = -(c8 * z + c1 * cos(theta) * C + C*A + c2 * sin(theta) * C + C*B- C) / c5;

affX = x+c3*w1;
affY = y+c4*w2;
affZ = z+c5*w3;
x1 = affX./(affX+affY+affZ);
y1 = affY./(affX+affY+affZ);
chromX = x./(x+y+z);
chromY = y./(x+y+z);

figure(1);
subplot(3,2,1); plot(lambda,w1); axis tight; grid on; title('w1')
subplot(3,2,2); plot(lambda,w2); axis tight; grid on; title('w2')
subplot(3,2,3); plot(lambda,w3); axis tight; grid on; title('w3')
subplot(3,2,4); plot(lambda,affX); grid on; axis tight; title('affX')
subplot(3,2,5); plot(lambda,affY); grid on; axis tight; title('affY')
subplot(3,2,6); plot(lambda,affZ); grid on; axis tight; title('affZ')

figure(2);
subplot(2,2,1); plot(lambda,[coneLong coneMid coneShort]); axis tight; grid on; title('cmfs');
subplot(2,2,2); plot(x1,y1); axis tight; grid on; title('affine spectral locus');
subplot(2,2,3); plot(chromX,chromY); axis tight; grid on; title('spectral locus');
% subplot(2,2,4); plot3(w1,w2,w3); axis tight; grid on; title('w-curve'); xlabel('w1'); ylabel('w2');
subplot(2,2,4); plot(lambda,[affX affY affZ]); axis tight; grid on; title('affine cmfs'); xlabel('lambda');
```